\documentclass{aa}  
\usepackage{graphicx}
\usepackage{txfonts}
\begin{document}
   \title{The HI content of Early--Type Galaxies from the ALFALFA survey}
   \titlerunning{HI in Early Type Galaxies}

   \subtitle{I. Catalogued HI sources in the Virgo cluster}

   \author{S. di Serego Alighieri
          \inst{1},
		  G. Gavazzi
		  \inst{2},
		  C. Giovanardi
		  \inst{1},
		  R. Giovanelli
		  \inst{3},
		  M. Grossi
		  \inst{1},
		  M.P. Haynes
		  \inst{3},
		  B.R. Kent
		  \inst{3},
		  R.A. Koopmann
		  \inst{4,5},
		  S. Pellegrini
		  \inst{6},
		  M. Scodeggio
		  \inst{7},
          \and
          G. Trinchieri\inst{8}
          }

   \offprints{S. di Serego Alighieri}

   \institute{INAF -- Osservatorio Astrofisico di Arcetri, Largo E. Fermi
   5, 50122 Firenze\\
              \email{sperello@arcetri.astro.it}
         \and
   Universit\`a di Milano--Bicocca, Piazza delle Scienze 3, 20126 Milano
   \and
   Center for Radiophysiscs and Space Research, Space Sciences Building, 
   Cornell University, Ithaca, NY 14853
   \and
   Department of Physiscs and Astronomy, Union College, Schenectady NY
   12308
   \and
   National Astronomy and Ionosphere Center \thanks{The National Astronomy
   and Ionosphere Center is operated by Cornell University under a cooperative 
   agreement with the National Science Foundation.}, Space Sciences
	  Building, Cornell University, Ithaca NY 14853
   \and
   Universit\`a di Bologna, Via Ranzani 1, 40127 Bologna
   \and
   INAF -- IASF Milano, Via Bassini 15, 20133 Milano
   \and
   INAF -- Osservatorio Astronomico di Brera, Via Brera 28, 20121 Milano
             }

   \date{Received: 3 July 2007; accepted: 8 September 2007 }
\authorrunning{S. di Serego Alighieri et al.}
 
  \abstract
{} 
   {We are using the Arecibo Legacy Fast ALFA survey (ALFALFA), which is
   covering 17\% of the sky at 21 cm, to
   study the HI content of Early-Type galaxies (ETG) in an
   unbiased way. The aim is to get an overall picture of the hot, warm
   and cold ISM of ETG, as a function of galaxy mass and environment, to
   understand its origin and fate,
   and to relate it to the formation and evolution history of these
   objects.
   }
   {This paper deals with the first part of our study, which is devoted to
   the 8-16 deg. declination strip in the Virgo cluster. In this sky region,
   using the Virgo Cluster Catalogue (VCC), we have defined an optical sample 
   of 939 ETG, 457 of which are brighter than the VCC completeness limit at
   $B_T=18.0$. We have correlated this optical sample
   with the catalogue of detected HI sources from ALFALFA.
   }
   {Out of the 389 ETG from the VCC with $B_T\le 18.0$, outside the 1 deg. 
   region of poor HI detection around M87, and corrected for background
   contamination of VCC galaxies without a known radial velocity, only 9 galaxies
   (2.3\%) are detected in HI with a completeness limit of 3.5 and 7.6$\times
   10^7 M_{\odot}$ of HI for dwarf and giant ETG, respectively. 
   In addition 4 VCC ETG with fainter magnitudes are
   also detected. Our HI detection rate is lower than previously claimed.
   The majority of the detected ETG appear to have peculiar morphology and
   to be located near the edges of the Virgo cluster.
   }
{Our preliminary conclusion is that cluster ETG contain very little 
neutral gas, with the exceptions of a few peculiar dwarf galaxies at the edge 
of the ETG classification and of very few larger ETG, where the cold gas could 
have a recent external origin.
}


   \maketitle
%

\section{Introduction}

Some Early--Type Galaxies (ETG), in particular the massive ellipticals, contain
large quantities of hot gas ($T\sim 10^7 K$), which can amount up to at least 
several $10^{10}M_{\odot}$ or about 1\% of the total stellar mass (see e.g. Mathews 
\& Brighenti 2003, for a review). The presence of warm gas ($T\sim 10^4 K$)
in ETG has been studied with several emission line surveys (e.g. Trinchieri \&
di Serego Alighieri 1991, Goodfrooij et al. 1994, Sarzi et al. 2006), which
have detected up to  about $10^5M_{\odot}$ of ionized gas. Cold gas ($T\sim 10^2
K$) has been detected in a number of ETG from HI 21cm observations (e.g. Knapp
et al. 1985, Huchtmeier 1994, Morganti et al. 2006), which find about 
$10^8M_{\odot}$ of neutral hydrogen in some objects. However an unbiased survey of
the HI content of ETG is still lacking, mainly because pointed observations
have preferentially selected ETG, where the presence of cold gas was likely,
and blind HI surveys were not deep enough or did not cover a large enough solid
angle in the sky.
In fact previous estimates of the HI detection rate for ETG, i.e. of the 
percentage of ETG detected in HI, vary between 15\% (Knapp et al. 1985 and 
Conselice et al. 2003) and more than 50\% (Morganti et al. 2006, see also 
Table 4 in Bregman et al. 1992).

The interest for studying the content of HI in ETG is based mainly on the
possibility that it can give important clues on the formation and evolution 
of this type of galaxies, also in relation with the other gas phases. Some
current models for ETG galaxy formation call for an important role of galaxy
merging, which is likely to happen also quite late, when most of the stars have
already formed. The presence of cold gas in an ETG would be a sign of
possible merging, particularly if it has disturbed morphology and/or
kinematics. Furthermore D'Ercole et al. (2000) investigated the effect
of tidal encounters on hot gas flows, and found the creation of dense,
cold filaments. However, in a previous set of simulations 
D'Ercole \& Ciotti (1998)
have studied hot gas flows in quite flattened galaxies (S0 or flat
Ellipticals), and found that cold filaments are also
created at the interface between inflowing and outflowing gas.

The Arecibo Legacy Fast ALFA survey (ALFALFA, Giovanelli et al. 2005 and 2007) is
covering in an unbiased way a large area of the sky (7074 square degrees) and
is detecting HI sources down to a relatively faint integrated flux density:
\begin{equation}
F_{lim} = 4.46\times 10^{-3} \times S/N \times \sqrt{W50} \times \sigma_{rms} Jy~km~s^{-1}
\end{equation}
where W50 is the velocity width in $km~s^{-1}$ of the line
profile at 50\% of the peak, $\sigma_{rms}$ is the r.m.s. noise in mJy
at $10~km~s^{-1}$ resolution and S/N is the required signal--to--noise
ratio. Since in ALFALFA, on average, $\sigma_{rms}=2.0 mJy$, then for S/N=6.5 and
$W50=100~km~s^{-1}$ the limiting flux density corresponds to $3.8\times 10^7 
M_{\odot}$ of neutral hydrogen at the distance of the Virgo cluster,
assumed to be 16.7 Mpc. 
ALFALFA therefore gives an unprecedented opportunity for studying
the cold gas content of ETG in an unbiased way. The strategy we have adopted 
to exploit this opportunity is briefly the following. We start by selecting a
sample of ETG, which should be as complete as possible
down to a given galaxy mass or luminosity, and should cover in an uniform way both the
cluster and the field environments, since these are likely to influence the gas
content in different ways. We then search for HI in these galaxies using the
ALFALFA survey data in a two--step approach. The first step is a simple 
cross--correlation between our ETG sample and the catalogue of detected HI sources
produced by the ALFALFA team (e.g. Giovanelli et al. 2007). The second step
involves searching in the ALFALFA datacubes for HI detections at the positions
-- and velocities, when available -- of the ETG in our sample. Hopefully
this second step can detect sources at a lower S/N ratio than the blind search used by the
ALFALFA team to produce their catalogue of detected HI sources. Furthermore
the second step can give us a better local estimate of the survey noise at the
position of each ETG, and therefore a better estimate of the upper
limits for those galaxies where no HI is detected.

In this first paper we report on the results from the first step on the Virgo
cluster, for the declination strip between 8 and 16 degrees. Our study is 
complemetary to the study of the dwarf elliptical galaxies, which
Koopmann et al. (in preparation) are conducting among the detected ALFALFA HI
sources.

\section{Selection of the ETG sample and HI detections}

The existence of the Virgo Cluster Catalogue (VCC, Binggeli et al. 1985, and the 
revision in Binggeli et al. 1993) makes
the selection of a sample of ETG in this cluster relatively
straightforward. The VCC is essentially complete, independently of morphological
type, down to $B_T \sim 18$, which corresponds to $M_B \sim -13.1$ for the
assumed distance to the Virgo cluster. However the VCC assigns a multiple 
morphological type to some of the galaxies, making it difficult to construct
a well defined subsample of ETG. We have therefore decided to adopt the
morphological classification univoquely given to VCC galaxies by GOLDMine
(Gavazzi et al. 2003). In the uncertain VCC cases GOLDMine has decided the
classification based on more recent information or on visual inspection of
more recent images.
We include in our ETG sample all VCC galaxies for which GOLDMine
gives a classification type between -3 and 1 (i.e. dS0, dE/dS0, dE(d:E),
E-E/S0, S0). We exclude galaxies which have $cz > 3000 km~s^{-1}$, since they
are not members of the Virgo cluster. Finally we limit our analysis to the
declination strip from 8.0 to 16.0 degrees, which is the only one for which
the catalogue of detected HI sources from ALFALFA is available (Giovanelli et
al. 2007 and private communication). This strip contains the largest fraction
of the Virgo cluster area covered by the VCC and more than 70\% of its
galaxies, centered around M87.
Our VCC ETG sample has 939 galaxies, 457 of which are brighter than the
completeness limit of $B_T = 18.0$.

\begin{figure}
\centering
\includegraphics[width=9.0cm]{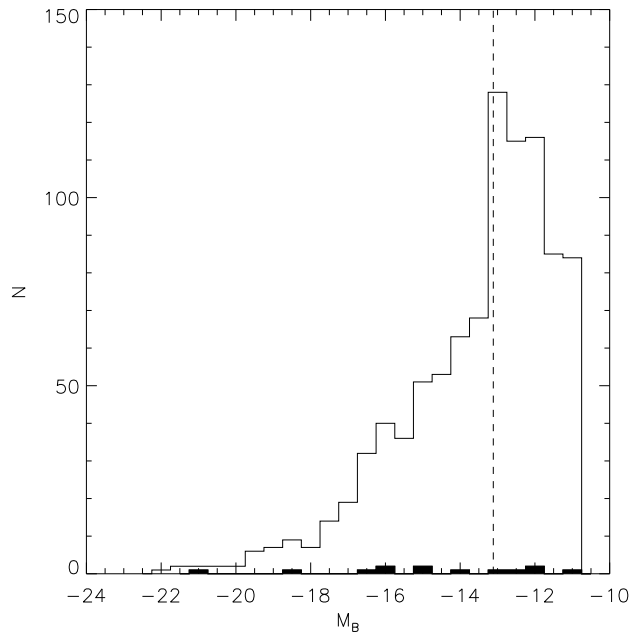}
\caption{The absolute magnitude distribution of the ETG detected in HI
(filled
histogram) with respect to all the ETG in the VCC (empty histogram). The
dashed vertical line corresponds to the VCC completeness limit at $B_T =
18.0$.
}
\label{Fig1}
\end{figure}

We have then cross--correlated this input optical ETG sample with the catalogue
of HI detections by ALFALFA in the same sky area (Giovanelli et 
al. 2007 and private communication). We have used a search radius of 180 arcsec, 
large enough to avoid possible biases from the uncertainties in the
optical and HI positions, and also from the large angular extent of a few 
galaxies in the catalog.
We obtain a total of 13 VCC ETG detected in HI, as listed in Table I. 
ALFALFA HI detections come at different signal levels (see Table 1): 
code 1 are detections typically with $S/N \ge 6.5$; code 2 are detections
at lower S/N (to $\sim 4.5$), which coincide spatially with an optical object
of known similar redshift; code 4 are possible detections with still lower S/N. 

%
\begin{table*}
\begin{minipage}[t]{\columnwidth}
\caption{VCC Early Type Galaxies detected in ALFALFA}
\label{Table I}
\centering
\renewcommand{\footnoterule}{}  
\begin{tabular}{lllrccrrcclc}
\hline \hline
ID& Other & $B_T$ & Type\footnote{GOLDMine type: -3=dS0 -2=dE$/$dS0 -1=dE(d:E) 0=E-E$/$S0 
1=S0} & Type & Opt. Pos. & $cz_{opt.}$ & $cz_{HI}$ & $M_{HI}$ & Code\footnote{See the text for
explanations} & $M_B$ & log($M_{HI}/L_B$) \\
~ & Name &~ & GM & VCC & RA Dec & $km~s^{-1}$ & $km~s^{-1}$ & $10^7 M_\odot$ &~ &~ & $M_\odot/L_\odot$ \\
\hline
VCC 21   & IC 3025  & 14.75 & -3 & dS0(4)       & 121023.0+101118 &  486 &  485 &  5.3 & 2 & -16.36 & -1.0 \\
VCC 93   & IC 3052  & 16.3  & -1 & dE2          & 121348.1+124126 &  910 &  841 &  3.5 & 1 & -14.8  & -0.6 \\
VCC 209  & IC 3096  & 15.15 & -3 & dS0?         & 121652.4+143055 & 1208 & 1263 &  3.6 & 1 & -15.96 & -1.0 \\
VCC 304  &          & 16.3  & -1 & dE1 pec?     & 121843.8+122308 &  155 &  132 &  3.2 & 1 & -14.8  & -0.7 \\
VCC 355  & NGC 4262 & 12.41 &  1 & SB0          & 121930.6+145238 & 1359 & 1367 & 49.1 & 1 & -18.70 & -1.0 \\
VCC 421  &          & 17.0  & -1 & dE2          & 122030.7+133109 &      & 2098 &  3.4 & 2 & -14.1  & -0.3 \\
VCC 881  & NGC 4406 & 10.06 &  0 & S0$_1$(3)/E3 & 122611.8+125646 & -244 & -302 &  8.0 & 2 & -21.05 & -2.7 \\
VCC 956  &          & 18.75 & -1 & dE1,N:       & 122656.4+125741 &      & 2151 &  9.2 & 1 & -12.36 &  0.8 \\
VCC 1142 &          & 19.0  & -1 & dE           & 122855.2+084855 &      & 1306 &  4.7 & 1 & -12.1  &  0.6 \\
VCC 1202 &          & 20.0  & -1 & dE?          & 122933.6+131146 &      & 1215 & 14.5 & 1 & -11.1  &  1.5 \\
VCC 1964 &          & 18.0  & -1 & dE4:         & 124319.7+085710 &      & 1495 &  4.3 & 4 & -13.1  &  0.3 \\
VCC 1993 &          & 15.3  &  0 & E0           & 124412.0+125631 &  845 &  925 &  4.5 & 2 & -15.8  & -0.9 \\
VCC 2062 &          & 19.0  & -1 & dE:          & 124759.9+105815 & 1146 & 1141 & 32.7 & 1 & -12.1  &  1.5 \\
\hline
\end{tabular}
\end{minipage}
\end{table*}

The ALFALFA HI detection threshold and completeness limit depend on 
the velocity width (Eq. 1). The average velocity width (W50) for the
HI detected ETG is 88 and 235 $km~s^{-1}$ for the 10 dwarfs and for the 
3 other ETG respectively. Since ALFALFA is complete to better than 90\%
at S/N = 6.5 for dwarfs and at S/N = 8.5 for giants (Saintonge 2007), 
these completeness limits correspond to 3.5 and $7.6\times 10^7 M_\odot$ of HI.

Only 9 of the 457 ETG in our sample
brighter than the VCC completeness limit (at $B_T = 18.0$)
are detected in HI, which corresponds to a percentage
of 2.0\%, considerably lower than previous estimates.
Even if we exclude the 59 ETG with $B_T\le 18.0$ contained in the one degree 
region around M87 (see Fig. 2), where 
the strong radio continuum source reduces considerably  the ALFALFA HI detection
sensitivity (Giovanelli et al. 2007), and the 11 ETG out of the 133
without a measured radial velocity (i.e. the same percentage of the ETG
with measured radial velocity which have $cz > 3000~km~s^{-1}$), which are
probably background galaxies,
the detection rate increases only slightly, to 2.3\%, 
which is still much lower than previous estimates.
In addition we have estimated the possible contamination on the complete VCC
sample due to field galaxies within 3000 $km~s^{-1}$, based on the
luminosity density and function given by Binggeli (1988) for the cluster and
the field. We derive an average value for this contamination of 1.4\%, not
depending strongly on the apparent magnitude limit. Its effect on the
detection rate is negligible.
Furthermore Binggeli et al. (1985) with the updates in Binggeli et al.
(1987) consider uncertain the cluster membership for 28 galaxies out of the
457 ETG in our sample brighter than the completeness limit of $B_T = 18.0$.
Some of these are likely background galaxies, with a percentage which is
then smaller or equal to 6.1\%.

Figure 1 shows the distribution of absolute B-band magnitudes for the
ETG in our whole sample, and for those that are detected in HI.
Considering the accuracy allowed by poissonian statistics for the small number 
of detected galaxies, there is no evidence that the detection rate depends on 
the galaxy luminosity: for example, 2 out of 55 ETG brighter than $M_B = -17.0$
($3.6\pm 2.5\%$) and 7 ETG out of 402 with $-17.0 < M_B \le -13.1$ ($1.7\pm
0.7\%$) are detected in HI. The ratio between the amount of neutral hydrogen 
and the B-band luminosity varies by about 4 orders of magnitude among
the ETG detected in HI (see the last column of Table 1 and Figure 2).
The strong correlation between the $M_{HI}/L_B$ ratio and the B-band luminosity
is mostly induced by the ALFALFA HI detection limits. In particular optically faint
ETG with a small $M_{HI}/L_B$ ratio would not be detected by ALFALFA.
However we remark on the
absence of bright ETG with a high $M_{HI}/L_B$ ratio, similar to that of the dwarf
ETG, which would have been easily seen, if they existed.

\begin{figure}
\centering
\includegraphics[width=9.0cm]{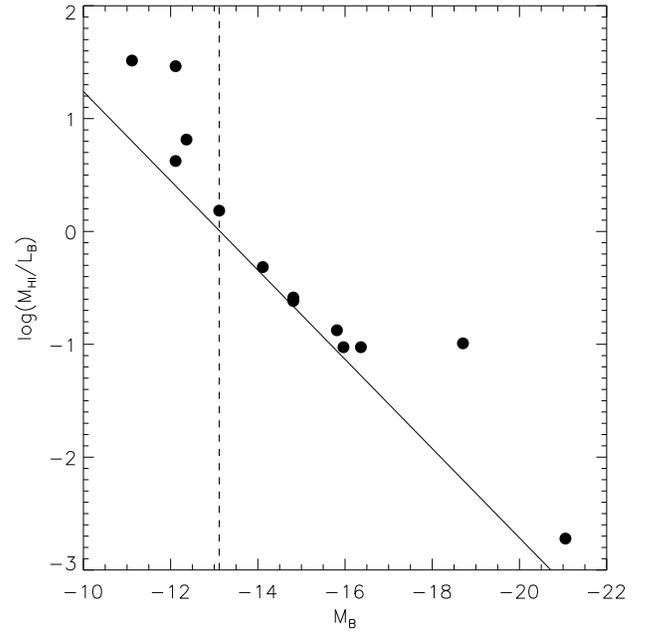}
\caption{The dependence of the ratio between the neutral hydrogen mass and the
B--band luminosity with the B--band absolute magnitude for the 13 detected ETG. 
The dashed line is the VCC completeness limit at $B_T=18.0$, and the continuous 
line represents a detection limit of $3\times 10^7 M_{\odot}$ of HI.
}
\label{Fig2}
\end{figure}

We also notice the large fraction of peculiar (or morphologically
uncertain) ETG among
those with HI (see the VCC type in Table 1). One can argue that the presence 
of peculiarities enhances the chances that an ETG contains neutral hydrogen
(van Gorkom \& Schiminovich 1997) and that bona fide 
elliptical and lenticular galaxies would have an HI detection rate even
lower than reported here. In fact several of the detected galaxies are at the
border of the ETG classification.

Figure 3 shows the position in the sky of the ETG detected in HI. Even
considering the lower HI detection sensitivity around M87 (the dark grey
circle in Fig. 3), it is apparent that the ETG detected in HI are
preferentially at the edges of the Virgo cluster and in the region of the
so--called cloud M in the North--Western side of the Virgo
cluster (Ftaclas, Fanelli \& Struble 1984), where the X-ray surface brightness is low
(B\"ohringer et al. 1994) and spiral galaxies appear to be less
HI--deficient
than in the rest of the cluster (Gavazzi
et al. 1999). An important result is also that 809 ETG of our sample in
the Virgo cluster (391 with $B_T\le 18.0$) have been well
observed by ALFALFA (they are out of the 1 deg. region around M87) and
have not been detected, i.e. they
contain less neutral hydrogen than the ALFALFA completeness limit of
3.5 and $7.6\times 10^7 M_{\odot}$ for dwarf ETG and for the other ETG,
respectively.

\begin{figure}
\centering
\includegraphics[width=9.0cm]{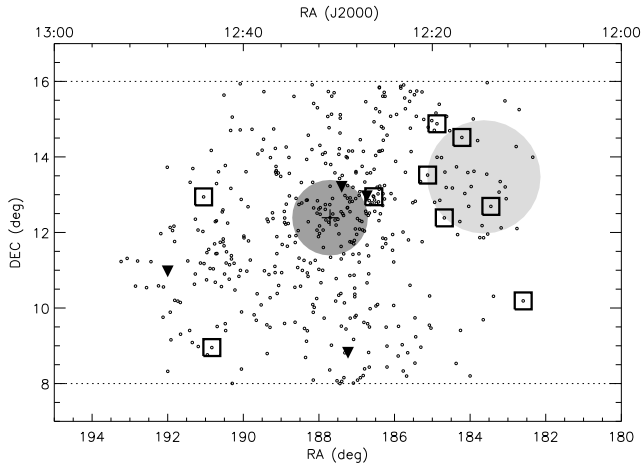}
\caption{Position of the Early Type galaxies in the 8-16 deg.
declination strip in the Virgo cluster: the dots are the ETG in the VCC
with $B_T\le 18.0$; squares and triangles are the ETG detected in HI, with 
$B_T\le 18.0$ and with $B_T > 18.0$, respectively.
The cross marks the position of
the central cD galaxy (M87), and the dark grey circle is the 1 deg. radius
zone where the HI detection is disturbed by M87 (Giovanelli et al. 2007).
The light grey circle shows the position of the M cloud (Ftaclas, Fanelli \&
Struble 1984).
}
\label{Fig3}
\end{figure}

\section{Notes on individual objects and comparison with previous work}

The dE4 galaxy VCC 748 has a tentative HI detection (code=4) in the ALFALFA
catalogue. However
it has been reobserved at 21cm as part of the follow--up programme and not detected to a 
$\sigma_{rms} = 1.46~mJy$ (Kent et al. in preparation). Using S/N=6.5 and 
assuming a velocity width of 88 km/s (the average for the detected dwarf ETG) 
this corresponds to an upper limit to the 21cm flux density of 0.4 $Jy~km~s^{-1}$ 
and to an upper limit to the HI mass at the distance of the Virgo cluster
of $2.6\times 10^7 M_{\odot}$. The other ETG detected with code=4 (VCC 1964) has 
not yet been reobserved, and its detection should be taken as only tentative.

We note that the most optically
luminous detected galaxy (VCC 881, also NGC 4406 and M86) is also the
ETG with the largest H$\alpha$ luminosity of the sample of ETG with
hot gas studied in H$\alpha$ by Trinchieri \& di Serego Alighieri (1991). The ionized
gas in this object has a complex structure of filaments and arcs,
possible sign of a recent merging event.

We have carried out an analysis of the previous HI detections
in the VCC area, not detected in the ALFALFA
catalogue, in order to understand the possible origin of the discrepancy
between our HI detection rate for ETG and those previously claimed.
The GOLDMine compilation reports HI detections for 11 ETG of our extended sample
of 939 galaxies. Only two of these (VCC 355 and VCC 2062) are present in Table I, 
i.e. are also detected by ALFALFA. For these two galaxies the previous 
observations (Burstein et al. 1987, and Hoffman et al. 1993) agree reasonably 
well with the ALFALFA measurements. Of the remaining 9 ETG, only two, VCC 1630 and 
VCC 575, have a previous HI detection at a very low level, below the ALFALFA 
detection limit (Lake \& Schommer 1984). For all the other 7 ETG (VCC 608, VCC 
763, VCC 1226, VCC 1619, VCC 1949, VCC 2012 and VCC2095) a careful inspection of 
the literature reveals the inconsistency of the previous HI detections. In
particular, VCC 608 (NGC 4323) was observed by Huchtmeier \& Richter (1986),
who state in a note that all the HI flux in the area comes from NGC 4321.
VCC 763 (NGC 4374) was observed by Davies \& Lewis (1973), who probably had
baseline problems. The best upper limit for the 21cm flux of VCC 763 is 7.5 $Jy~km~s^{-1}$ by
Heckman et al. (1983). VCC 1226 (NGC 4472) was tentatively detected at
2$\sigma$ by Davies \& Lewis (1974), but the best upper limit for the 21cm
flux is 0.10 $Jy~km~s^{-1}$ by Bregman et al. (1988). Peterson (1979)
claimed a detection for VCC 1619 (NGC 4550), but this galaxy was not
detected in HI by Krumm \& Salpeter (1979, $S_{21cm} < 2.1 Jy~km~s^{-1}$) nor by
Duprie \& Schneider (1996, $S_{21cm} < 1.3 Jy~km~s^{-1}$). 
Huchtmeier \& Richter (1986)
claimed a detection for VCC 1949 (NGC 4640), but nothing is visible in
their spectrum, and the galaxy was not detected by Haynes \& Giovanelli (1986,
$S_{21cm} < 0.9 Jy~km~s^{-1}$). VCC 2012 was detected by Huchtmeier \&
Richter (1986) but with the incredible velocity width of 472$km~s^{-1}$ for
a dE3 galaxy. The low level detection of VCC 2095 (NGC 4762) by Krumm \&
Salpeter (1976) was later disclaimed, while Giovanardi et al. (1983) give
an upper limit for the 21cm flux of 0.5 $Jy~km~s^{-1}$.
Therefore there is no real discrepancy between the previous HI measurements
and the results of ALFALFA, as far as the VCC ETG in our declination strip 
are concerned.

In order to understand the reasons for our much lower detection rate, 
we analyse in more detail the recent HI study of Morganti et al. 
(2006), who have surveyed the HI content of the elliptical and lenticular
galaxies of the SAURON sample. The SAURON team (de Zeeuw et al. 2002) has
selected from the Lyon-Meudon Extragalactic Database (LEDA, Paturel et al.
1997) a sample of 29 E and 51
S0 galaxies in clusters, and 47 E and 86 S0 galaxies in the field with the
following criteria: $cz\le 3000 km~s^{-1}$, $M_B\le -18.0$, $-6.0^{\circ}\le 
Dec.\le 64^{\circ}$. 
Out of these 213 galaxies they have selected a representative sample of 12 E 
and 12 S0 galaxies in clusters, and 12 E and 12 S0 galaxies in the field.
They show that the representative sample of 48 galaxies has no bias in the
$M_B$ vs. ellipticity plane with respect to the total sample of 213 galaxies.
Morganti et al. (2006) have concentrated on the SAURON subsample of 24 galaxies 
in the field and observed all those with declination larger than 23 degrees,
as well accessible from Westerbork. They have however excluded NGC 3032, which is a
SAB0, and included NGC 4150 and NGC 4278, which belong to the cluster SAURON
sample. They have then observed at 21cm 4 E and 8 S0 galaxies. They
detect HI in 4 E and in 5 S0 galaxies, with a detection rate of 75\%. 
Their pointed 21 cm observations are
deeper than the ALFALFA survey. Still, at the ALFALFA detection level, 
3 E and 3 S0 are detected out of the 12
galaxies observed, i.e. a detection rate of 50\%.
We have however to consider that the SAURON sample is selected at brighter
magnitudes ($M_B\le -18.0$) than ours ($M_B\le -13.11$). Reducing our samples
to the same magnitude limit as SAURON, ALFALFA detects HI in 2 ETG out of 32
(outside the 1 deg. region around M87), a
detection rate which is still 8 times lower than the one of Morganti et al.
(2006). Part of this difference might be due to the well--known higher HI 
detection rate in the field than in the clusters (e.g. Solanes et al. 2001, and
references therein).

\section{Conclusions}

We are exploiting the great opportunity given by the ALFALFA survey to
study in an unbiased way the HI content of ETG. The cross--correlation
of the ETG in the 8--16 deg. declination strip of the VCC and the catalogue
of ALFALFA detected sources for the 8--16 deg. declination strip on the
Virgo cluster has resulted in 13 ETG detected in HI, with a survey completeness
limit of 3.5 and $7.6\times 10^7 M_{\odot}$ of HI for dwarf and giant ETG, respectively. 
Only two of these ETG were already known to
contain neutral hydrogen. The HI detection rate for ETG down to
the VCC completeness limit of $B_T = 18.0$ varies between 2.3 and 2.0\%,
depending on whether or not one considers the region of lower HI detection
sensitivity around M87 and the correction for background galaxies. This
detection rate is considerably lower than previous claims, based on
incomplete surveys, and there is no clear evidence that it 
depends on the galaxy luminosity. The HI detected galaxies tend to
lie at the edges of the cluster and to have a peculiar morphology. 

Within the detection limits of ALFALFA, which affect differently large and
dwarf ETG, our preliminary conclusion is that cluster ETG contain very little or no 
neutral gas, with
the exception of some peculiar dwarfs, which are at the edge of the ETG
morphological classification, and of very few larger galaxies for which
the gas has a recent external origin. Given the very few large ETG with
HI, we suggest that the ETG merging rate with gas reach objects should be very
low in a cluster like Virgo.

We are currently improving the HI detection for the ETG of
our optical VCC sample by a search in the ALFALFA datacubes at the
optical positions and velocities, when available, which will be the
subject of a future paper (Grossi et al., in preparation). 
We are also planning to investigate the HI
content of a complete sample of ETG in the field.

\begin{acknowledgements}
We thank Bruno Binggeli for his kind advice on the VCC galaxies, and the referee
for useful comments. RG, MPH and BRK acknowledge partial support from NAIC, from 
NSF grants AST--0307661, AST--0435697 and AST--0607007, and from the Brinson 
Foundation. RAK acknowledges partial support from NAIC, from NSF grant
AST--0407011, and the hospitality of 
the Cornell University Department of Astronomy during a sabbatic visit.
\end{acknowledgements}

\end{document}